\documentclass[twocolumn]{2026SCGE_REVISED}
\usepackage{bm}
\usepackage{algorithm}
\usepackage{graphicx}
\usepackage{amsmath,amssymb,amsxtra,amsfonts}
\usepackage[switch,mathlines]{lineno}
\usepackage{hyperref}
\hypersetup{
  colorlinks=true,
  linkcolor=blue,
  citecolor=blue,
  urlcolor=blue,
  pdfborder={0 0 0}
}

\usepackage{xcolor}

\begin{document}
\ensubject{astrophysics}

\ArticleType{Article}
\SpecialTopic{SPECIAL TOPIC: }
\Year{2026}
\Month{January}
\Vol{69}
\No{1}
\DOI{??}
\ArtNo{000000}
\ReceiveDate{January 11, 2023}
\AcceptDate{April 6, 2023}

\title{Pandora cluster Lensing, AGN, and Transient Exploration (PLATE) from JWST Multi-Epoch Imaging. I. Discovery of a type II supernova candidate in a spiral galaxy at $z=0.7$}{Pandora cluster Lensing, AGN, and Transient Exploration (PLATE) from JWST Multi-Epoch Imaging. I. Discovery of a type II supernova candidate in a spiral galaxy at $z=0.7$}



\author[0009-0005-3823-9302]{Yuxuan Pang}
\affiliation{School of Astronomy and Space Science, University of Chinese Academy of Sciences (UCAS), Beijing 100049, China}
\correspondingauthor{Yuxuan Pang}
\email{pangyuxuan@ucas.ac.cn}

\author[0000-0002-9373-3865]{Xin Wang}
\affiliation{School of Astronomy and Space Science, University of Chinese Academy of Sciences (UCAS), Beijing 100049, China}
\affiliation{National Astronomical Observatories, Chinese Academy of Sciences, Beijing 100101, China}
\affiliation{Institute for Frontiers in Astronomy and Astrophysics, Beijing Normal University, Beijing 102206, China}
\correspondingauthor{Xin Wang}
\email{xwang@ucas.ac.cn}

\author[0000-0003-0853-6427]{Ping Chen}
\affiliation{Institute for Advanced Study in Physics, Zhejiang University, Hangzhou 310058, China}
\affiliation{Institute for Astronomy, School of Physics, Zhejiang University, Hangzhou 310058, China}

\author[0000-0002-1027-0990]{Subo Dong}
\affiliation{Department of Astronomy, School of Physics, Peking University, 5 Yiheyuan Road, Haidian District, Beijing 100871, China}
\affiliation{Kavli Institute of Astronomy and Astrophysics, Peking University, 5 Yiheyuan Road, Haidian District, Beijing 100871, China}
\affiliation{National Astronomical Observatories, Chinese Academy of Sciences, Beijing 100101, China}

\author[0009-0004-7133-9375]{Hang Zhou}
\affil{School of Astronomy and Space Science, University of Chinese Academy of Sciences (UCAS), Beijing 100049, China}

\author[0009-0007-6655-366X]{Shengzhe Wang}
\affiliation{National Astronomical Observatories, Chinese Academy of Sciences, Beijing 100101, China}
\affiliation{School of Astronomy and Space Science, University of Chinese Academy of Sciences (UCAS), Beijing 100049, China}

\author[0009-0006-1255-9567]{Qianqiao Zhou}
\affil{School of Astronomy and Space Science, University of Chinese Academy of Sciences (UCAS), Beijing 100049, China}

\author[0009-0005-8170-5153]{Xunda Sun}
\affil{School of Astronomy and Space Science, University of Chinese Academy of Sciences (UCAS), Beijing 100049, China}

\author{Jifeng Liu}
\affiliation{National Astronomical Observatories, Chinese Academy of Sciences, Beijing 100101, China}
\affiliation{School of Astronomy and Space Science, University of Chinese Academy of Sciences (UCAS), Beijing 100049, China}
\affiliation{Institute for Frontiers in Astronomy and Astrophysics, Beijing Normal University, Beijing 102206, China}

\author[0000-0003-1718-6481]{Hu Zhan}
\affiliation{National Astronomical Observatories, Chinese Academy of Sciences, Beijing 100101, China}
\affiliation{Kavli Institute of Astronomy and Astrophysics, Peking University, 5 Yiheyuan Road, Haidian District, Beijing 100871, China}

\author[0000-0002-3254-9044]{Karl Glazebrook}
\affiliation{Centre for Astrophysics and Supercomputing, Swinburne University of Technology, PO Box 218, Hawthorn, VIC 3122, Australia}

\author[0000-0002-2057-5376]{Ivo Labbé}
\affiliation{Centre for Astrophysics and Supercomputing, Swinburne University of Technology, PO Box 218, Hawthorn, VIC 3122, Australia}

\author[0000-0003-2804-0648]{Themiya Nanayakkara}
\affiliation{Sydney Institute for Astronomy, School of Physics, The University of Sydney, Sydney, NSW 2006, Australia}

\author[0000-0003-4263-2228]{David A. Coulter}
\affiliation{Space Telescope Science Institute (STScI), 3700 San Martin Drive, Baltimore, MD 21218, USA}

\author[0000-0002-2361-7201]{Justin D. R. Pierel}
\affiliation{Space Telescope Science Institute (STScI), 3700 San Martin Drive, Baltimore, MD 21218, USA}

\author[0000-0002-4410-5387]{Armin Rest}
\affiliation{Space Telescope Science Institute (STScI), 3700 San Martin Drive, Baltimore, MD 21218, USA}
\affiliation{Physics and Astronomy Department, Johns Hopkins University, Baltimore, MD 21218, USA}

\author[0000-0002-8296-2590]{Jujia Zhang}
\affiliation{Yunnan Observatories, Chinese Academy of Sciences, Kunming 650216, China}
\affiliation{International Centre of Supernovae, Yunnan Key Laboratory, Kunming 650216, China}
\affiliation{Key Laboratory for the Structure and Evolution of Celestial Objects, Chinese Academy of Sciences, Kunming 650011, China}


\AuthorMark{Pang Y.}

\AuthorCitation{Pang Y., et al.}

\abstract{We report the discovery and multi-wavelength analysis of a $z\sim0.7$ transient PLATE-23a in the Abell 2744 field, as the first results of the Pandora Lensing, AGN, and Transient Exploration (PLATE) project. Using multi-epoch JWST NIRCam imaging spanning 2022–2025, we detect PLATE-23a in 12 filters. Difference-imaging analysis reveals its rising and declining phases. The host galaxy of PLATE-23a is a barred spiral at $z=0.688$ with a stellar mass of $\sim 10^{10.4}M_{\odot}$ and a star formation rate of $\sim5.1M_{\odot}~\rm yr^{-1}$, placing it on the star-forming main sequence. Bayesian light-curve classification strongly favors a type IIP supernova (SN) origin. It lies $\sim 13\rm kpc$ (after lensing correction) from the galaxy center in a region of low local star formation, suggesting the progenitor may have migrated from a distant star-forming clump. Physical properties derived from blackbody modeling indicate a temperature decreasing from $\sim8010\rm K$ to $\sim6100\rm K$ at around 65 days after the explosion; the late-time SED is consistent with entering the radioactive decay phase at about one rest-frame year. This work demonstrates the power of deep, multi-epoch JWST observations for studying transients at cosmological distances.}


\maketitle
\Authorfootnote

\section{Introduction} \label{sec:intro}
With its deeper sensitivity, higher spatial resolution, and unique infrared wavelength coverage, the James Webb Space Telescope (JWST) has opened new opportunities for studying the faint ($m_{\rm AB}\sim 30\rm mag$) variable Universe. After several years of science operations, JWST has accumulated multiple epochs of observations in several well-studied fields, leading to significant advances in the study of active galactic nuclei (AGNs), transients, and related phenomena \cite{2025A&A...704L...8L, 2025ApJ...990..149K, 2025ApJ...979..250D, 2025arXiv251012572P, 2026ApJ...999..152S, 2026ApJ..1002..162F}.

Among those regions, Abell 2744 (the Pandora Cluster) is one of the most extensively observed extragalactic fields. As a merging galaxy cluster \cite{2024ApJ...974...92B}, its mass distribution and lensing properties are well characterized after more than a decade of detailed modeling \cite{2011MNRAS.417..333M, 2014MNRAS.444..268R, 2015ApJ...811...29W, 2014MNRAS.443.1549J, 2016ApJ...819..114K, 2023ApJ...952...84B, 2023A&A...670A..60B}, revealing a large lensing area with magnifications exceeding 2. Previously, the main core of A2744 was targeted by the Hubble Frontier Fields program, which provided a wealth of imaging and spectroscopic data, including HST and Spitzer imaging \cite{2017ApJ...837...97L,2020ApJS..247...64S}, HST grism spectroscopy \cite{2015ApJ...812..114T}, MUSE integral-field spectroscopy \cite{2021A&A...646A..83R}, and ALMA observations \cite{2022ApJS..263...38K}. These observations yield multi-band spectral energy distributions (SEDs), photometric redshifts, and spectroscopic redshifts for a large number of sources.

In the JWST era, Abell 2744 was one of the first fields observed (the GLASS Early Release Science program \cite{2022ApJ...935..110T}). Since then, this field has been imaged with NIRCam in every JWST cycle, including prominent programs such as UNCOVER \cite{2024ApJS..270....7W}, MAGNIF \cite{2023jwst.prop.2883S}, All the Little Things  (ALT) \cite{2024arXiv241001874N}, and SAPPHIRES \cite{2025arXiv250315587S}. These observations span from 2022 June to 2025 October with a cadence of several months, and additional data will continue to be released in future cycles. These conditions make Abell 2744 an ideal field for detecting transients and variability. Motivated by this, we started the Pandora Lensing, AGN, and Transient Exploration (PLATE) project, which aims to search for such sources and study their properties.

Supernovae (SNe) are among the most important classes of transients. High-redshift Type Ia SNe can serve as potential probes of dark energy \cite{2022ApJ...938..110B}, while the properties of core-collapse SNe provide constraints on the stellar initial mass function and the cosmic star formation rate density (e.g., Strolger et al. \cite{2015ApJ...813...93S}). 
Type II SNe, on the other hand, are core-collapse explosions of massive stars that retain hydrogen-rich envelopes at the time of explosion. Historically, they were divided into SNe IIP, which exhibit a roughly 100-day phase of nearly constant optical luminosity, and SNe IIL, whose light curves decline more rapidly and approximately linearly in magnitude after peak luminosity \cite{1979A&A....72..287B, 1989ApJ...342L..79Y, 1994A&A...282..731P, 2002AJ....123..745R}. In the classical picture, the plateau of an SN IIP is produced as the shock-heated ejecta expand and cool. Hydrogen recombination maintains the photospheric temperature near the recombination temperature, while the recombination front recedes through a relatively massive hydrogen-rich envelope, causing the photospheric radius and optical luminosity to evolve slowly \cite{1993ApJ...414..712P, 2009ApJ...703.2205K}. The more rapid evolution traditionally associated with SNe IIL is generally attributed to a lower-mass hydrogen-rich envelope, potentially resulting from enhanced stellar winds or binary interaction, which produces a shorter and less prominent recombination phase.
Over the past decade, HST has discovered SNe at $z\lesssim2$ in various deep fields \cite{2011ApJS..197...35G, 2014AJ....148...13R, 2012ApJS..199...25P, 2014ApJ...783...28G, 2024ApJS..272...19O}, and large ground-based surveys have identified numerous superluminous SNe ($M_{\rm AB}\approx-21 \rm mag$) \cite{2018ApJ...854...37S, 2019ApJS..241...16M, 2019ApJS..241...17C, 2017MNRAS.470.4241P}. More recently, JWST observations have led to significant progress in SN science, including using time delays between multiple lensed images of a SN to constrain the Hubble constant \cite{2024ApJ...961..171F, 2023ApJS..269...43Y, 2024ApJ...967L..37P}, discovering high-redshift SNe \cite{Siebert_2024, 2026ApJ..1002..162F, Siebert_2026, Coulter_2026, coulter2026, 2026arXiv260524088A}, and building SN catalogs with preliminary classifications from optical to mid-infrared light curves \cite{2025ApJ...979..250D, 2026ApJ...998..115Y}. These results open a new era for SN studies.
 
In this paper, we present the first PLATE result: PLATE-23a, a supernova at redshift $z=0.688$ with extensive multi-epoch, multi-filter observations. The paper is organized as follows. Section \ref{sec:data} describes the multi-epoch NIRCam observations of Abell 2744; then outlines the image reduction process, the difference-imaging strategy, and the selection and photometry of the SN object. In Section \ref{sec:result}, we discuss the host galaxy properties and the classification of PLATE-23a. Section \ref{sec:Summary} summarizes our conclusions. Throughout this work, we express magnitudes using the AB system \cite{1983ApJ...266..713O} and assume a $\Lambda$CDM cosmology with parameters $\Omega_m$ = 0.3, $\Omega_{\Lambda}$  = 0.7 and $H_0$ = 70 km $\text{s}^{-1}$ $\text{Mpc}^{-1}$.

\section{Data and Reduction} \label{sec:data}

\subsection{NIRCam Imaging Data} \label{subsec:data}
We retrieved all publicly available NIRCam imaging (stage 2, \texttt{cal.fits} files) of the Abell 2744 field by querying the Mikulski Archive for Space Telescopes (MAST\footnote{\url{https://mast.stsci.edu/}}) through \texttt{astroquery}. To recover the full SEDs of variable sources over multiple epochs, we focused on 12 JWST filters with observational coverage spanning more than one month: F070W, F090W, F115W, F140M, F150W, and F200W in the short-wavelength (SW) channel, and F277W, F356W, F410M, F430M, F444W, and F480M in the long-wavelength (LW) channel. The search was centered at (R.A., decl.) = (3.57, –30.375) with a radius of 6'. The resulting data includes observations from various JWST programs, including JWST-GLASS \cite{2022ApJ...935..110T}, UNCOVER \cite{2024ApJS..270....7W}, ALT \cite{2024arXiv241001874N}, and SAPPHIRES \cite{2025arXiv250315587S}. It's worth to mention that some of the medium-band filters were excluded in doing the time domain analysis since it is only observed in only a single epoch (e.g., F162M from the MegaScience program; \citealt{2024ApJ...976..101S}). In total, we obtained about 10,000 NIRCam exposures spanning from June 2022 to Nov 2025; a summary of the programs, filters, and observation epochs is listed in Table \ref{table:obs}.

\begin{table*}[t]
\footnotesize
\caption{Information of the NIRCam image observation.}
\label{table:obs}
\tabcolsep 5pt 
\begin{tabular*}{\textwidth}{ccccc}
 \toprule
 Proposal ID & Filter & Obs Date & Ref\\
    \hline
 1176 & F090W, F115W, F150W, F200W, F277W, F356W, F410M, F444W & 2023-08 & \cite{2023AJ....165...13W}\\
    1324 & F090W, F115W, F150W, F200W, F277W, F356W, F444W & 2022-06, 2023-07 & \cite{2022ApJ...935..110T}\\
    2561 & F090W, F115W, F150W, F200W, F277W, F356W, F410M, F444W, F480M & 2022-11, 2023-07, 2023-08, 2024-07 & \cite{2024ApJS..270....7W} \\
    2756 & F115W, F150W, F200W, F277W, F356W, F444W & 2022-10, 2022-12 & \cite{2023Natur.618..480R}\\
    2883 & F480M & 2023-11 & \cite{2023jwst.prop.2883S}\\
    3073 & F115W, F150W, F200W, F277W, F356W, F444W & 2023-10, 2024-07 & \cite{2023jwst.prop.3073C} \\
    3516 & F070W, F090W, F356W & 2023-11, 2023-12 & \cite{2024arXiv241001874N}\\
    3538 & F410M & 2023-10, 2023-11 & \cite{2023jwst.prop.3538I} \\
    3990 & F090W, F115W, F140M, F200W, F277W, F356W, F430M, F444W, F480M & 2023-10, 2024-08, 2024-10 & \cite{2025ApJ...983..152M}\\
    4111 & F070W, F090W, F140M, F430M, F480M & 2023-11 & \cite{2024ApJ...976..101S} \\
    6434 & F070W, F090W, F150W, F200W, F356W, F444W & 2024-10, 2024-11, 2024-12, 2025-06, 2025-10 & \cite{2025arXiv250315587S} \\
 \bottomrule
\end{tabular*}
\end{table*}

\subsection{Image Reduction and Alignment}\label{subsec:imgalign}
Spatially resolved difference imaging is a powerful tool for identifying and classifying transients that relies on two key techniques: point-spread function (PSF) template modeling across different observations and precise astrometric alignment of the exposures. In ground-based data, the PSF is dominated by seeing and can be well described by Gaussian-like functions; for space-based telescopes, however, the non-Gaussian PSF features and field rotation make the problem significantly more challenging \cite{2024AJ....168..174B}. Moreover, optimal difference imaging requires sub-pixel alignment accuracy, whereas the standard JWST pipeline is typically only good to about 1 pixel \cite{bushouse_2026_20058613,2025ApJ...979..250D,2025ApJ...981L...9P}.
Therefore, in this work, we adopt the data-processing framework used in several recent transient studies \cite{2025ApJ...979..250D,2025ApJ...981L...9P,2026ApJ..1002..162F,2026arXiv260524088A}, employing JWST/HST Alignment Tool (JHAT, \citet{armin_rest_2023_7892935}) \cite{armin_rest_2023_7892935} for image alignment, \texttt{HOTPANTS} (Becker 2015, \cite{2015ascl.soft04004B}) for PSF modeling and difference-image construction, and \texttt{space-phot} \cite{pierel_2024_12100100} for photometry. In this Section, we describe the detailed steps we take to identify variable candidates in the F444W band and to perform photometry for PLATE-23a.

We started with all stage 2 data (\texttt{cal.fits} files) retrieved from the MAST portal. These files were processed by the JWST pipeline version 1.20.2 (Bushouse et al. 2025\cite{bushouse_2025_17515973}) with the Calibration Reference Data System (CRDS) pmap-1464, which provides improved suppression of 1/f noise, background subtraction, and artifact removal. 

We then performed further astrometric alignment on the stage 2 data using the \texttt{JHAT}, which is designed to register individual NIRCam exposures to a reference catalog. In this work, we constructed the reference catalog by combining Gaia Early Data Release 3 (EDR3, \cite{2021A&A...649A...1G}) with the UNCOVER DR3 catalog \cite{2024ApJ...976..101S}. Finally, we combined the astrometry-aligned images into science mosaics (in different epochs, separation method in Section \ref{subsec:diffimg}) using stage 3 of the JWST pipeline (v1.20.2), adopting the native pixel scales of $0.''03$ for SW and $0.''06$ for LW filters. With this procedure, the median absolute astrometric offset between the observations and the reference catalog is $0.''013$, with a standard scatter of $0.''006$. Our alignment precision is slightly lower than recent cosmos-web results\cite{2026ApJ..1002..162F}, mainly due to the bright galaxies and the intracluster light in the foreground galaxy clusters.

\subsection{Difference Imaging}\label{subsec:diffimg}
As shown in Table \ref{table:obs}, the Abell 2744 field benefits from extensive multi-epoch observations. To begin the transient search, we grouped all observations into 50-day-wide epochs in each band. We then selected the F444W filter — the reddest broad LW band, which offers the highest probability of detecting high-redshift transients — to test the difference-imaging procedure. In total, the F444W imaging spans eight epochs (Jun. 2022, Oct. 2022, Jul. 2023, Oct. 2023, Jul. 2024, Oct. 2024, Dec. 2024, and Oct. 2025, respectively).

After grouping observations into epochs and getting the science mosaics (Section \ref{subsec:imgalign}), we generated difference images for all combinations of the eight epochs. We used \texttt{HOTPANTS} to match the spatially varying PSF kernel between different epochs within the F444W filter, adopting the general parameter setup and convolution direction described in previous work \cite{2014ApJ...795...44R,2025ApJS..280...29A,2026ApJ..1002..162F}. For each pair of epoch images (\texttt{IMG1} and \texttt{IMG2}), we produced both \texttt{IMG1-IMG2} and \texttt{IMG2-IMG1} and chose the one with the lower median background as the final difference image. This approach accounts for the fact that some epochs have shorter exposure times, which can result in larger uncertainty of the \texttt{HOTPANTS} PSF models. The total overlapping area covered by the F444W imaging is $\sim55~\rm arcmin^2$.

\subsection{Variable Source Identification}
The final step is to build the transient catalog from the difference images across all epochs. Since we are preparing another paper presenting all variable sources detected in the F444W band in the Abell 2744 field, along with their general properties and classification (Pang et al. in prep), we provide only a brief summary of the selection process here.

We first ran the Python wrapper \texttt{sewpy}\footnote{\url{https://github.com/megalut/sewpy}} for \texttt{SExtractor}\cite{1996A&AS..117..393B} on the difference images in all filters to identify both positive and negative detections exceeding 5$\sigma$ of the local background. Since most false detections arise from image artifacts in the cores of bright cluster galaxies, we imposed an additional constraint requiring no other detection within 8 pixels ($\sim 0''.5$) of any candidate. This criterion may reject some dual AGNs or SNe on the outskirts of AGN host galaxies, but given the limited survey volume, the number of such missed cases is expected to be negligible \cite{2025MNRAS.536.3016P, 2020MNRAS.495.3252S}.

With these criteria, the sample of potential transients or intrinsically variable sources was reduced to a manageable number of candidates in the F444W difference images (on the order of thousands). We then visually inspected each candidate to assess whether it exhibited intrinsic variability (e.g., AGN-like activity with a fluctuating light curve near the galaxy center) or transient features (e.g., supernova-like outbursts at off-center locations) across epochs. 
We inspected multi-band image cutouts, F444W difference image cutouts, multi-band light curves and properties of host galaxies (when available).
We ultimately selected about 70 variable sources, including transients, AGNs, and possible lensed stars. The most common false positives were hot pixels near image edges, diffraction spikes of bright stars, and cosmic rays.

\subsection{PLATE-23a}

\begin{figure}[H]
\centering
\includegraphics[width=0.48\textwidth]{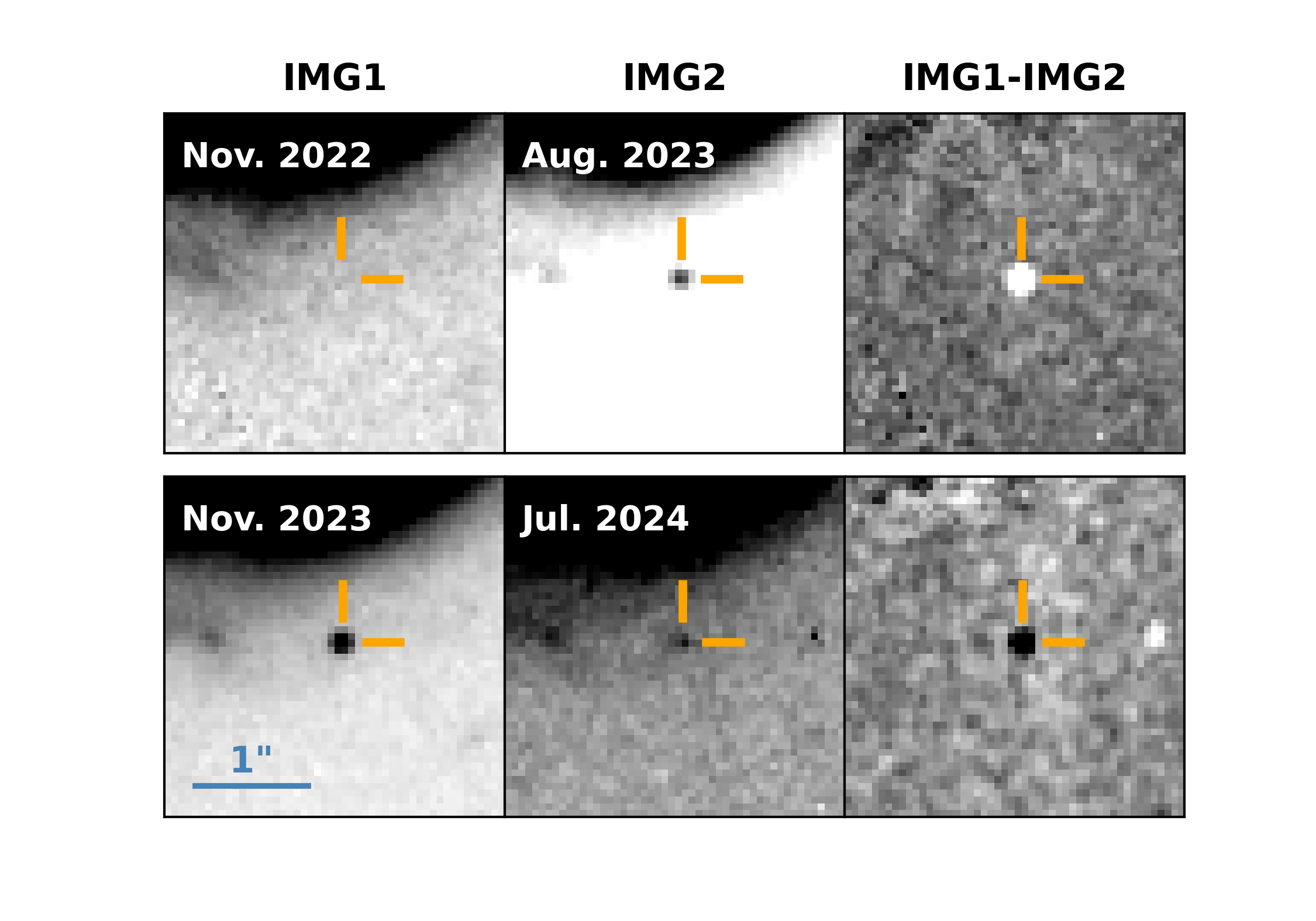}
\caption{Difference imaging of the PLATE-23a in the F444W filter, produced with \texttt{HOTPANTS}. From left to right, the original cutout images at two different epochs and the \texttt{HOTPANTS} resulting difference image. The top and bottom rows illustrate the brightening and fading phases of the SN, respectively. Owing to the large photometric variation, both processes are clearly identified in the difference image. The position of the SN and a 1'' scale bar are marked in the figure.}
\label{fig:diff-img}
\end{figure}

\begin{figure*}[h!]
\centering
\includegraphics[width=0.95\textwidth]{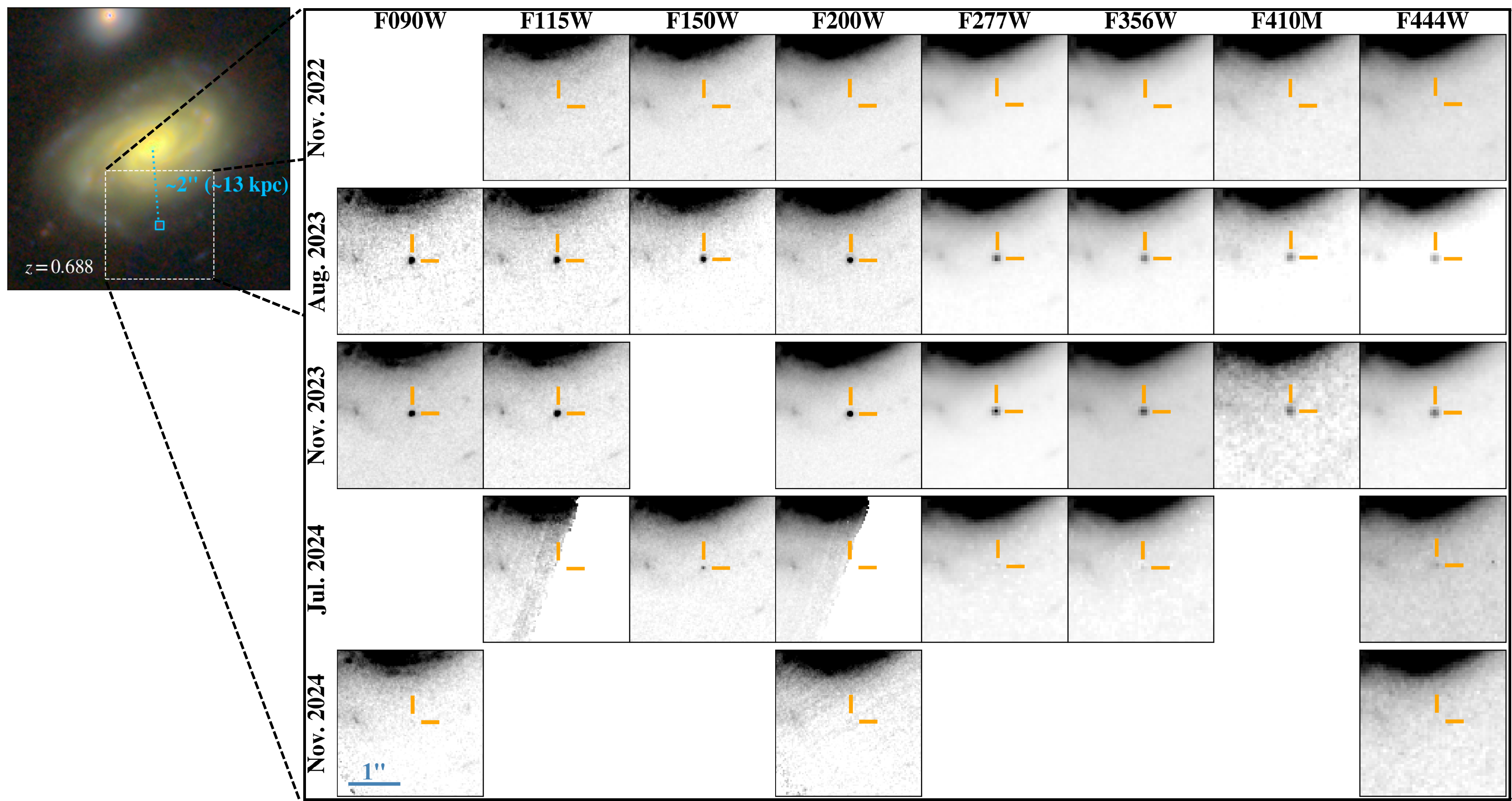}
\caption{Multi-band and multi-epoch overview of PLATE-23a. Upper left panel: RGB composite image (by the science mosaic from all epochs, R: F444W, G: F277W, B: F115W) showing the SN and its distance to the center of the host galaxy. Right panels: Zoom-in NIRCam imaging of the SN, arranged with different filters in columns (labeled at the top) and five epochs in rows (labeled on the left).}
\label{fig:multi-ph}
\end{figure*}

PLATE-23a (R.A. = 03h33m09.9216s and decl. = -30d24m01.2229s) is the most striking transient which has the brightest detections and the largest number of detection epochs and bands among all the variable sources. It is clearly detected in the difference images during both the rising and declining phases of its light curve as shown in Figure \ref{fig:diff-img}. Multi-band inspection shows that the SN is captured in 12 NIRCam filters across 5 epochs, as shown in Figure \ref{fig:multi-ph}, making it an excellent SN candidate for detailed classification and properties study.

To obtain the multi-band photometric light curves of PLATE-23a, we performed aperture photometry with a radius of $0''.18$, chosen to match the typical FWHM of the longest-wavelength F480M filter (from \citet{2024ApJ...962..139Z}), on the science mosaics in each epoch and filter using the \texttt{space-phot} package. The measured total fluxes, originally in units of $\rm MJy~sr^{-1}$, were converted to AB magnitudes adopting the pixel scales of the images ($0''.03~\rm pixel^{-1}$ for SW and $0''.06~\rm pixel^{-1}$ for LW). We note that systematic zero-point uncertainties are negligible compared to the aperture photometry uncertainties \cite{2022RNAAS...6..191B}. The multi-band photometry of PLATE-23a is listed in Table \ref{table:photometry}.

\begin{table}[t]
\footnotesize
\caption{Multi-band photometry of PLATE-23a.}
\label{table:photometry}
\tabcolsep 12pt 
\begin{tabular*}{0.4\textwidth}{cccc}
 \toprule
 Epoch & MJD & Filter & AB Mag (err)\\
    \hline
  1 & 59891  &  F115W  &  29.42(u) \\
  1 & 59887  &  F150W  &  29.65(u) \\
  1 & 59891  &  F200W  &  29.69(u) \\
  1 & 59891  &  F277W  &  29.70(u) \\
  1 & 59891  &  F356W  &  29.68(u) \\
  1 & 59890  &  F410M  &  29.08(u) \\
  1 & 59891  &  F444W  &  29.33(u) \\
    \hline
  2 & 60142  &  F090W  &  25.39(0.02) \\
  2 & 60148  &  F115W  &  25.25(0.02) \\
  2 & 60158  &  F150W  &  25.48(0.02) \\
  2 & 60158  &  F200W  &  25.64(0.02) \\
  2 & 60153  &  F277W  &  26.02(0.02) \\
  2 & 60153  &  F356W  &  26.12(0.02) \\
  2 & 60158  &  F410M  &  26.52(0.05) \\
  2 & 60145  &  F444W  &  26.54(0.04) \\
    \hline
  3 & 60277  &  F070W  &  27.27(0.05) \\
  3 & 60259  &  F090W  &  26.00(0.01) \\
  3 & 60242  &  F115W  &  25.42(0.01) \\
  3 & 60262  &  F140M  &  25.47(0.02) \\
  3 & 60246  &  F200W  &  25.58(0.01) \\
  3 & 60242  &  F277W  &  25.90(0.01) \\
  3 & 60243  &  F356W  &  26.01(0.01) \\
  3 & 60248  &  F410M  &  26.30(0.08) \\
  3 & 60262  &  F430M  &  26.37(0.07) \\
  3 & 60246  &  F444W  &  26.40(0.02) \\
  3 & 60260  &  F480M  &  26.74(0.13) \\
    \hline
  4 & 60494  &  F150W  &  27.71(0.07) \\
  4 & 60494  &  F277W  &  28.65(0.19) \\
  4 & 60494  &  F356W  &  28.48(0.13) \\
  4 & 60494  &  F444W  &  28.53(0.18) \\
    \hline
  5 & 60605  &  F090W  &  29.15(u) \\
  5 & 60607  &  F200W  &  28.40(u) \\
  5 & 60605  &  F444W  &  28.77(u) \\
 \bottomrule
\end{tabular*}
\begin{tablenotes}
\item[1] Letter ``u'' in magnitude error correspond to 3$\sigma$ upper limit results.
\end{tablenotes}
\end{table}

\subsection{Lens Models} \label{subsec: lensmodel}

In this work, we adopt the publicly available lens model of Bergamini et al. \citep{2023ApJ...952...84B}, constructed with the \texttt{LENSTOOL} software \citep{2007NJPh....9..447J} and constrained by new GLASS-JWST, UNCOVER, and MUSE observations. This model incorporates 149 multiple images, an increase of $\sim$66\% over previous models. At the redshift of PLATE-23a ($z=0.688$; Section \ref{subsec:host}), the magnification factor is $\mu=1.32$.

To evaluate the systematic uncertainty introduced by the choice of lens model, we also examined the earlier model of Bergamini et al. (2023b) \citep{2023A&A...670A..60B}, which yields $\mu=1.49$. We therefore adopt a fiducial magnification of 1.32 for the classification and SED fitting (Figures \ref{fig:classification}, \ref{fig:blackbody}, \ref{fig:model_lightcurve}), and propagate the difference between the two models as a systematic error of $±0.13\rm~mag$ when estimating luminosities (Figures \ref{fig:relation} and \ref{fig:LT_compare}).

\section{Results and Discussion} \label{sec:result}

\subsection{Redshift and Host Properties} \label{subsec:host}
The host galaxy of the SN appears to be a barred spiral with multiple spiral arms (see upper left RGB image in Figure \ref{fig:multi-ph}). Interestingly, the SN is located roughly 13 kpc ($\sim2''$) from the galaxy center, which is well beyond its effective radius ($0''.9-1''.2$, 5.6-7.4 kpc) from the single Sérsic \texttt{GalfitM} \cite{2010AJ....139.2097P, 2022A&A...664A..92H} fitting result (after lensing correction by model from \cite{2023ApJ...952...84B}). Owing to the depth of the imaging, however, it is difficult to determine whether the SN lies directly on a spiral arm. Previous optical spectroscopy ($3700–8800\AA$) obtained with the AAOmega multi-object spectrograph on the Anglo-Australian Telescope indicates a redshift of 0.688 (from Table 5 in \citet{2011ApJ...728...27O}, redshift quality $Q=4$). We note that the ALT Data Release 1 also provides grism redshifts from the F356W observations, reporting $z=0.674$ at the galaxy center and $z=0.685$ at a star-forming clump. These values are consistent with the AAOmega spectroscopic redshift, considering the systematic uncertainties in the grism redshifts introduced by the extended morphology of the host galaxy.

\begin{figure}[H]
\centering
\includegraphics[width=0.48\textwidth]{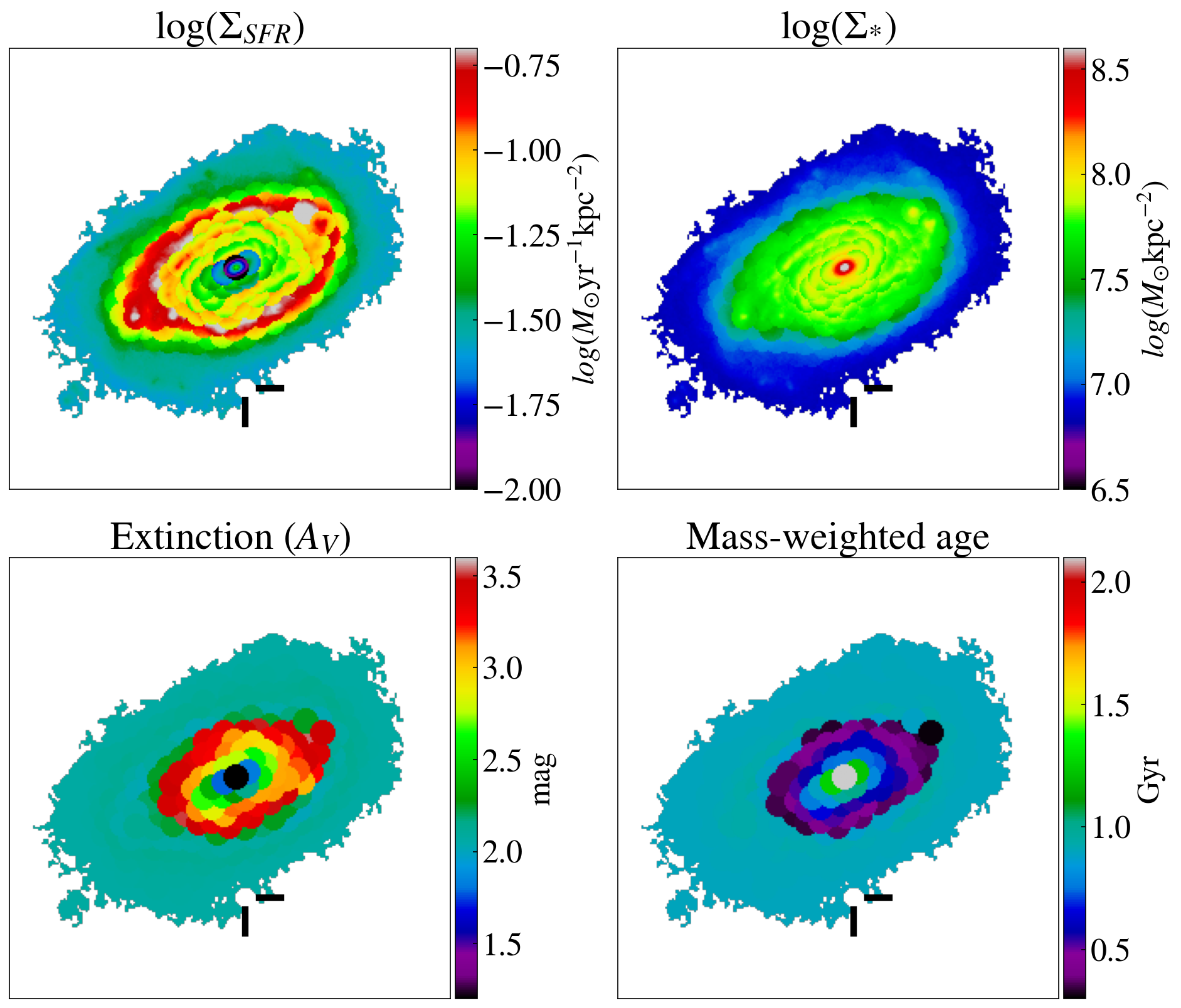}
\caption{Pixel-resolved properties of the host galaxy of PLATE-23a derived with the \texttt{piXedfit} procedure (by the science mosaic from all epochs). From left to right and top to bottom, the panels display the star formation rate surface density, stellar mass surface density, extinction, and mass-weighted stellar age. The pixel binning radius is about 0.23'' during the fitting. The location of PLATE-23a is highlighted in the panels by black lines. The region within the FWHM of the PSF centered on the SN has been masked to avoid contamination.}
\label{fig:host}
\end{figure}

At such a redshift, the JWST NIRCam data provide multi-band, high-resolution imaging. To further characterize the host galaxy, we performed the pixel-resolved SED fitting\footnote{\url{https://github.com/aabdurrouf/piXedfit}}. To maximize the SNR, we combined all available exposures in each filter into deep science mosaics, including other medium band data from MegaScience \cite{2024ApJ...976..101S}, using stage 3 of the JWST pipeline (v1.20.2), adopting a common pixel scale of $0.''06$. We then spatially binned the images with a minimum bin size of 5 pixels, requiring a total F444W SNR greater than 10 within each bin. To minimize contamination from the SN, we masked the region within the FWHM of the PSF centered on the SN position \cite{2024ApJ...962..139Z}. The spatially resolved SEDs were compared against a library of 10,000 template spectra generated from a delayed exponential star formation history (SFH) assuming Chabrier et al. (2003) initial mass function (IMF) \cite{2003PASP..115..763C} with a fixed redshift at 0.688, generated by \texttt{Pixedfit}. We include dust and nebular emission in the fitting. The binned pixel fitting results are presented in Figure \ref{fig:host}. 

The stellar mass surface density of the host galaxy decreases from the center outward, and its Sérsic index is about 1 by \texttt{GalfitM} (0.94 in F070W to 1.19 in F480M band), consistent with properties of nearby barred spirals. The total stellar mass is $10^{10.4\pm0.3} M_{\odot}$, placing it in the normal galaxy regime. The star formation rate (SFR) surface density increases with radius, consistent with the inside-out growth scenario for galaxies at $z\le1$. Such surface density is enhanced in the spiral arm regions, and the total SFR is $5.1\pm 0.1 M_{\odot} yr^{-1}$. Therefore, the host galaxy lies on the star-forming main sequence (SFMS, adopted from Speagle et al. 2014 \cite{2014ApJS..214...15S}) at redshift 0.688. The mass-weighted stellar age also correlates well with the SFR: regions of active star formation show younger stellar ages. Meanwhile, the extinction map reveals higher extinction in regions with higher SFR, indicating a good spatial correlation between star formation and the molecular gas mass traced by dust.

In the vicinity of the SN, the fitting results show a relatively low SFR and stellar mass surface density ($\Sigma_{*}=10^{6.8\pm0.1}M_{\odot}~\rm kpc^{-2}$, $\Sigma_{\rm SFR}=0.026\pm 0.004M_{\odot}~\rm kpc^{-2}~yr^{-1}$) with moderate extinction ($A_{\rm V}=1.9\pm 0.3\rm~mag$). Assuming a Chabrier IMF \cite{2003PASP..115..763C} spanning 0.07–100$M_{\odot}$ \cite{2014AJ....147...94D,2021A&A...645A.100S}, roughly 6,000 massive stars ($>8M_{\odot}$) have formed in the last 10 Myr within each square kiloparsec. Thus, despite its relatively low surface brightness, the SN
environment contains a substantial local population of possible
core-collapse progenitors.

We quantify the alternative possibility that the progenitor formed
in the nearest prominent star-forming clump and subsequently
migrated to the SN position. For two regions of comparable effective
area, we assign relative weights
\begin{equation}
 W_{\rm local}=\Sigma_{\rm SFR,local}, 
 W_{\rm migrated}=\Sigma_{\rm SFR,clump} \times f_{\rm HV},
\end{equation}
where $f_{\rm HV}$ is the probability that a massive star is ejected
with sufficient velocity to reach the observed position. The clump
is located at a lensing-corrected projected distance of approximately
7 kpc and has $\Sigma_{\rm SFR,clump}/\Sigma_{\rm SFR,local}\sim8$.
For an assumed progenitor age of 10 Myr, migration requires a projected velocity of at least $v_\perp\simeq700\,{\rm km\,s^{-1}}$ (the true three-dimensional velocity would be larger). Hyper-runaway stars with such velocities are expected to be extremely rare. Even a binary-population model optimized to produce fast ejections predicts a probability of only $1.77\times10^{-5}$ per progenitor system for $v_{\rm ej}\geq400\,{\rm km\,s^{-1}}$ \cite{2019A&A...624A..66R, 2020MNRAS.497.5344E}. Adopting this value as an upper limit, $f_{\rm HV}$ gives
\begin{equation}
 f_{\rm HV} = \frac{W_{\rm migrated}}{W_{\rm local}}<10^{-3},
\end{equation}
within this two-hypothesis model. Dynamical interactions involving a very massive star or an intermediate-mass black hole can in principle produce velocities near $700\,{\rm km\,s^{-1}}$, but require exceptional cluster conditions \cite{1988Natur.331..687H, 2009MNRAS.396..570G, 2014ApJ...785L..23Z}. We therefore conclude that PLATE-23a most probably formed locally rather than migrating from the nearby star-forming clump.


\subsection{Classification} \label{subsec:class}

Following the method implemented for JADES and COSMOS-Web transients \cite{2025ApJ...979..250D,2026ApJ..1002..162F}, we classified PLATE-23a using the \texttt{STARDUST2} Bayesian light-curve classification tool \cite{2014AJ....148...13R}, which is built on the \texttt{SNCosmo}\cite{barbary_2025_15019859} fitting framework. \texttt{STARDUST2} compares the observed photometry with a library of spectrophotometric time-series templates for different SN types and identifies the most likely class based on the $\chi^2$ value. The fitted templates include the SALT3-NIR model for Type Ia SNe \cite{2007A&A...466...11G, 2021ApJ...923..265K, 2022ApJ...939...11P}, and a set of 27 Type II and 15 Type Ib/c templates from the Sloan Digital Sky Survey \cite{2008AJ....135..338F, 2008AJ....135..348S, 2010ApJ...708..661D}, the Supernova Legacy Survey \cite{2006A&A...447...31A}, and the Carnegie Supernova Project \cite{2006PASP..118....2H, 2009ApJ...696..713S, Morrell_2011}, extended to the near-infrared by Pierel et al. (2018) \cite{2018PASP..130k4504P}, to represent core-collapse SNe. Since most of the SNe in the template library lack long-term photometric coverage of their late-time fading, we used the Ni–Co–Fe decay timescale to extrapolate the light curves out to 300 days in the rest frame \cite{1994ApJS...92..527N}.

A nested sampling algorithm \cite{2004AIPC..735..395S} is used to compute likelihoods over the SN model parameter space, incorporating priors on dust parameters as described by Rodney et al. (2014) \cite{2014AJ....148...13R}. We used all detections in the first- and second-epoch photometry of PLATE-23a for classification, comprising 12 data points across 7 NIRCam filters, since most SN models lack late-time observations in the rest-frame near-infrared.

\begin{figure*}[h]
\centering
\includegraphics[width=0.98\textwidth]{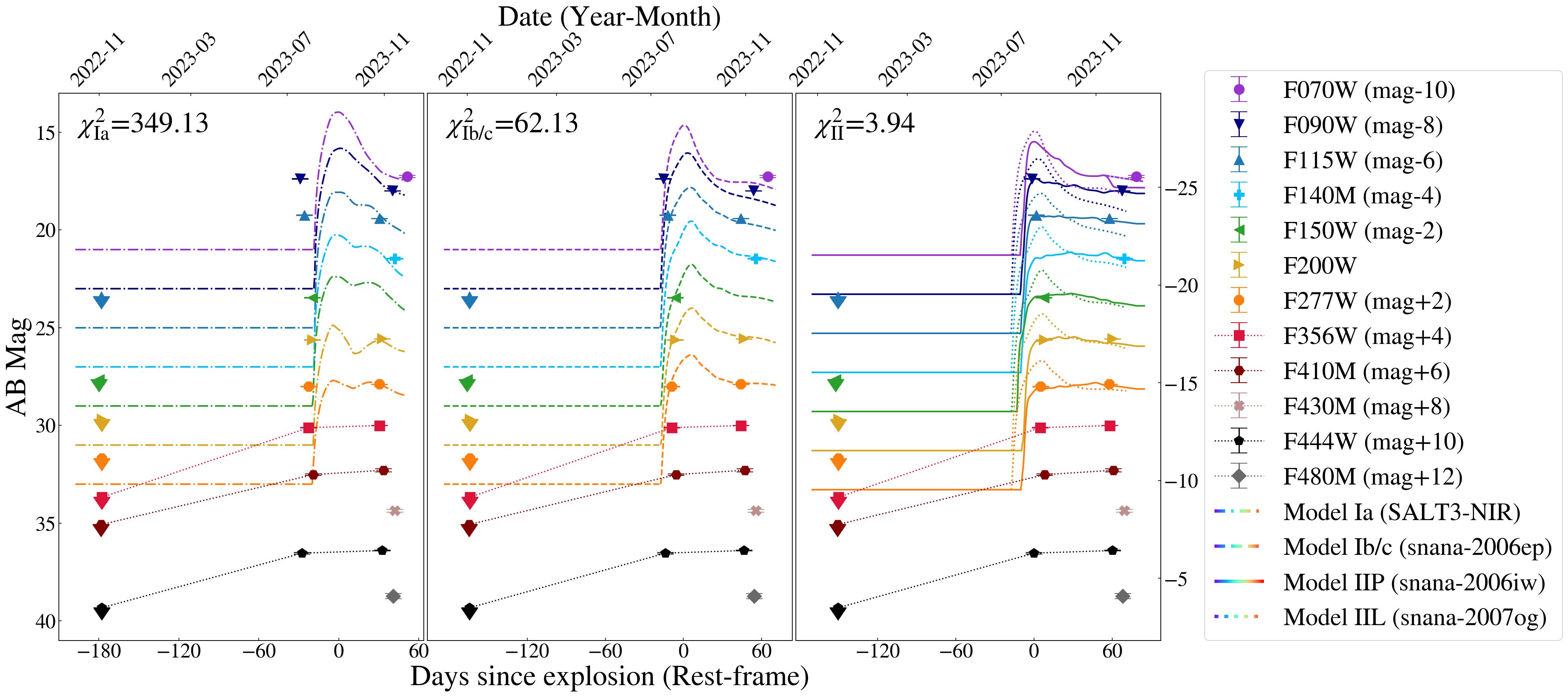}
\caption{Classification results based on the first two epochs of multi-band photometry of PLATE-23a. Multi-band photometry is shown as colored dots, with the best-fit models for Type Ia, Ib/c, and Type II SNe overlaid as dot-dashed, dashed, and solid lines in the three columns, respectively. The best-fitting template is the light curve of snana-2006iw, a Type II(P) SN. For comparison, the best-fit Type IIL SN model is also shown as dotted lines in the right panel. The dashed line in the F070W band in the right panel shows the light curve extrapolated beyond 55 rest-frame days using the $s_2$ plateau decline rate (see Section \ref{subsec:class}). The lower x-axis indicates rest-frame time since explosion, and the upper x-axis gives the corresponding JWST observation date. For clarity, the photometry in different wavebands has been vertically offset. Non-detections are plotted as 3$\sigma$ upper limits.}
\label{fig:classification}
\end{figure*}

Figure \ref{fig:classification} presents the best-fitting results of Type Ia, Ib/c, and II for PLATE-23a. Among all the SN models, snana-2006iw, a Type II(P) SN, provides the best fit, significantly outperforming the other models ($\Delta BIC <- 10$). We also tested the fitting procedure by leaving the redshift as a free parameter and found that snana-2006iw with a redshift around 0.7 still yields the best-fit results, corroborating that PLATE-23a is indeed associated with the galaxy at $z=0.688$ and is not a foreground or background interloper. The best-fit extinction at the host galaxy is $A_{\rm V}\sim 0.71\pm 0.02\rm~mag$ assuming the extinction curve from \citet{1999PASP..111...63F} with $R_{\rm V}=3.1$, which is smaller than the value derived from pixel-resolved SED fitting of the host galaxy at the location of PLATE-23a. The difference in extinction may be attributed to the different projected distances of the SN and the host galaxy. We adopt the best-fit extinction from the light-curve fitting to estimate the physical parameters of PLATE-23a discussed below. Table \ref{tab:fitting_re} shows the parameters of the best-fitting snana-2006iw model. 

We also examined the ALT slitless spectroscopy data at the SN position and found no significant residual features after subtracting the host-galaxy contamination. At the epoch of the ALT F356W grism observations (roughly 80 rest-frame days after explosion), the corresponding F356W photometry is about 26.2 mag, whereas the continuum sensitivity of the 70 ks F356W grism data is only about 25 mag; therefore, no continuum detection is expected. Moreover, the strongest emission line expected in the F356W band at late times is $Pa\alpha$, with an intrinsic luminosity of $10^{39-40}\rm erg~s^{-1}$ (i.e., \citet{2006MNRAS.368.1169P}). The line sensitivity of the ALT F356W grism data is also approximately $10^{39-40}\rm erg~s^{-1}$ at $z=0.688$. However, because the line is intrinsically broad (FWHM of several thousand $\rm km~s^{-1}$) and is further affected by residual host-galaxy contamination, the non-detection of emission-line features is also consistent with expectations.

\begin{table}[t]
\centering
\footnotesize
\caption{Best fitting parameters of snana-2006iw model}
\label{tab:fitting_re}
\tabcolsep 6pt 
\begin{tabular*}{0.48\textwidth}{cccc}
 \toprule
    Date & $t_0$ & Absolute Magnitude & $A_{V}$ \\
    MJD & $\rm erg~s^{-1}$ & V band \\
    \hline
    07-20-2023 & $60145.0\pm 0.6$ & $-17.4\pm0.1$ & $0.71\pm 0.02$ \\
 \bottomrule
\end{tabular*}
\begin{tablenotes}
\item[1] Using the lensing model from \cite{2023ApJ...952...84B}.
\end{tablenotes}
\end{table}

Type II SNe are core-collapse explosions of massive stars that retain hydrogen-rich envelopes at the time of explosion. Within the classical framework, the photometric diversity of Type II SNe is primarily associated with differences in the mass and structure of their hydrogen-rich envelopes. Nevertheless, the observed diversity can also be affected by the explosion energy, progenitor radius, $^{56}$Ni mass and mixing, ejecta-density structure, and interaction with circumstellar material \cite{2016MNRAS.455..423M, 2017ApJ...838...28M}. In the LOSS sample, SNe IIP and SNe IIL accounted for $69.9^{+5.1}_{-5.8}\%$ and $9.7^{+4.0}_{-3.2}\%$ of Type II events, respectively \cite{2011MNRAS.412.1441L}, although these fractions depend on the adopted classification criteria. More recent statistical studies indicate that plateau-related parameters, particularly the post-peak decline rate, form a continuous distribution rather than two clearly separated populations \cite{2014ApJ...786...67A, 2015ApJ...799..208S, 2016MNRAS.459.3939V, 2017AJ....154..211K}. It is therefore useful to characterize Type II light curves quantitatively rather than relying exclusively on the traditional IIP/IIL labels. In this framework, the post-peak evolution is divided into the initial rapid decline, the plateau or recombination-dominated phase, and the post-plateau radioactive tail; their average decline rates per 100 days are denoted by $s_1$, $s_2$, and $s_3$, respectively.

We noticed that the best-fit SNANA-2006iw model tends to under-predict the F070W flux after explosion. A closer inspection of the F070W light-curve reveals a suspicious drop around 60 rest-frame days. This drop is unlikely to represent the transition from the plateau to the radioactive tail, because such a transition would occur simultaneously across all bands (i.e., \citet{2019ApJ...873..127T}) and no corresponding flux decline is detected in any other filter. Further examination found that the template spectrum around rest-frame 4000 $\AA$ may suffer from data quality issues due to atmospheric absorption, appearing as a distinct dip. We therefore used the $s_2$ decline rate of SN 2006iw to extrapolate the F070W light curve beyond 55 days by an exponential decay, as shown by the dashed line in the right panel of Figure \ref{fig:classification}. The extrapolated light curve is in good agreement with the observed photometry. It is also worth noting that removing the F070W data point does not affect the final best-fit template or the classification result.

\begin{figure}[H]
\centering
\includegraphics[width=0.48\textwidth]{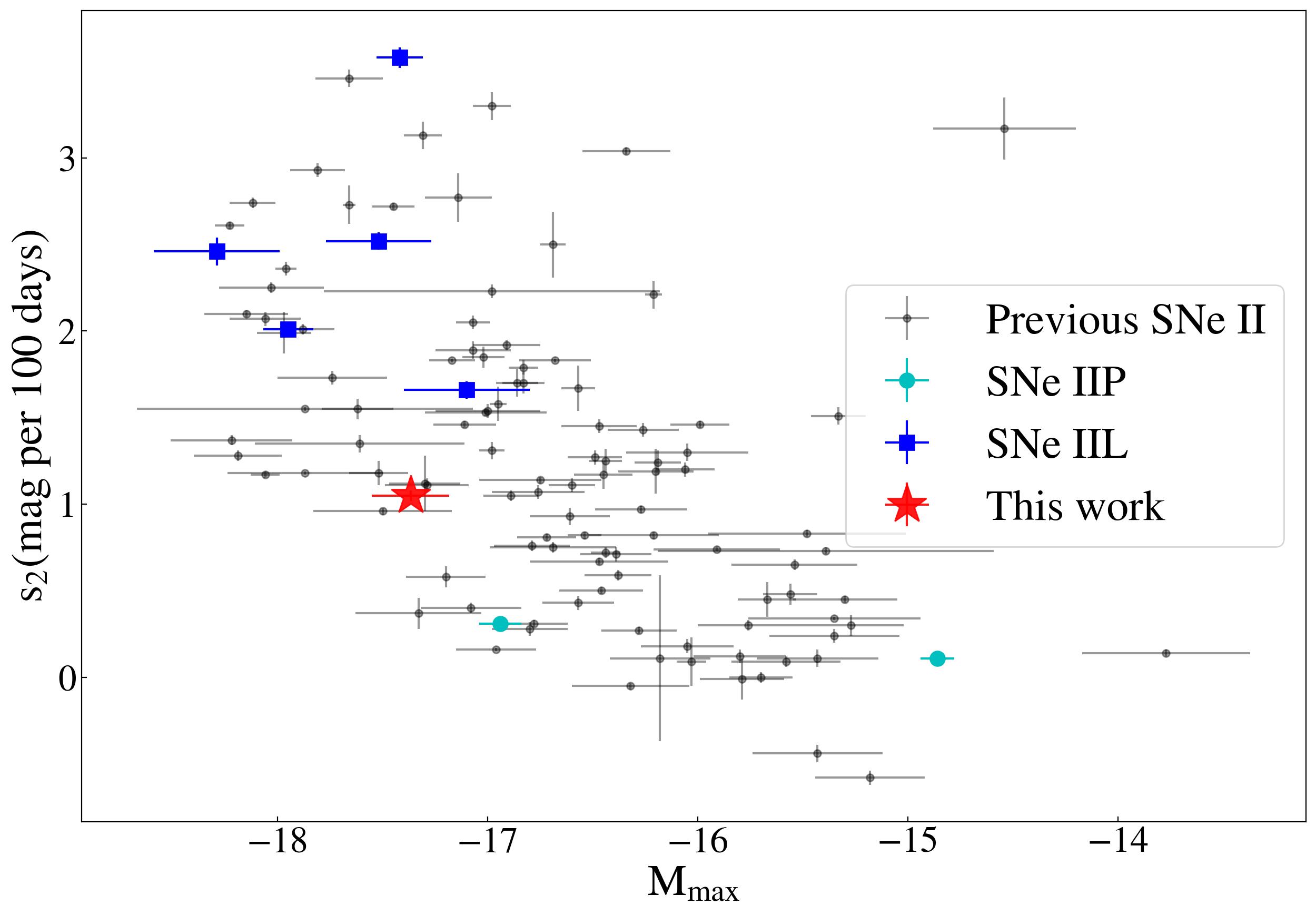}
\caption{Peak magnitude, Mmax (from the V band light-curve of the best fitting model), plotted against the decline rate during the “plateau” s2 of Type II SNe. Black dots represent local SNe from \citet{2014ApJ...786...67A}, including previously classified Type IIP (cyan dots) and Type IIL (blue squares) events. The PLATE-23a fitting result is shown in a red star.}
\label{fig:relation}
\end{figure}

Statistical analysis by \citet{2014ApJ...786...67A} found that the peak absolute magnitudes ($M_{\rm max}$) and the s2 parameter exhibit a trend wherein lower-luminosity SNe decline more slowly, which is consistent with the classic picture that SNe IIL are generally more luminous than SNe IIP. Figure \ref{fig:relation} shows the position of PLATE-23a in $M_{\rm max}$-$s_2$ parameter space from the best-fit model. PLATE-23a ($M_{\rm max}=-17.37$, $s_2=1.05$) also falls along this relation, indicating that the luminosity of this event doesn't deviate from the local Type II population.

\subsection{Properties of PLATE-23a} \label{subsec:properties}

Given the Type IIP classification, we estimate the basic physical properties of PLATE-23a in different phases.

\begin{figure}[H]
\centering
\includegraphics[width=0.48\textwidth]{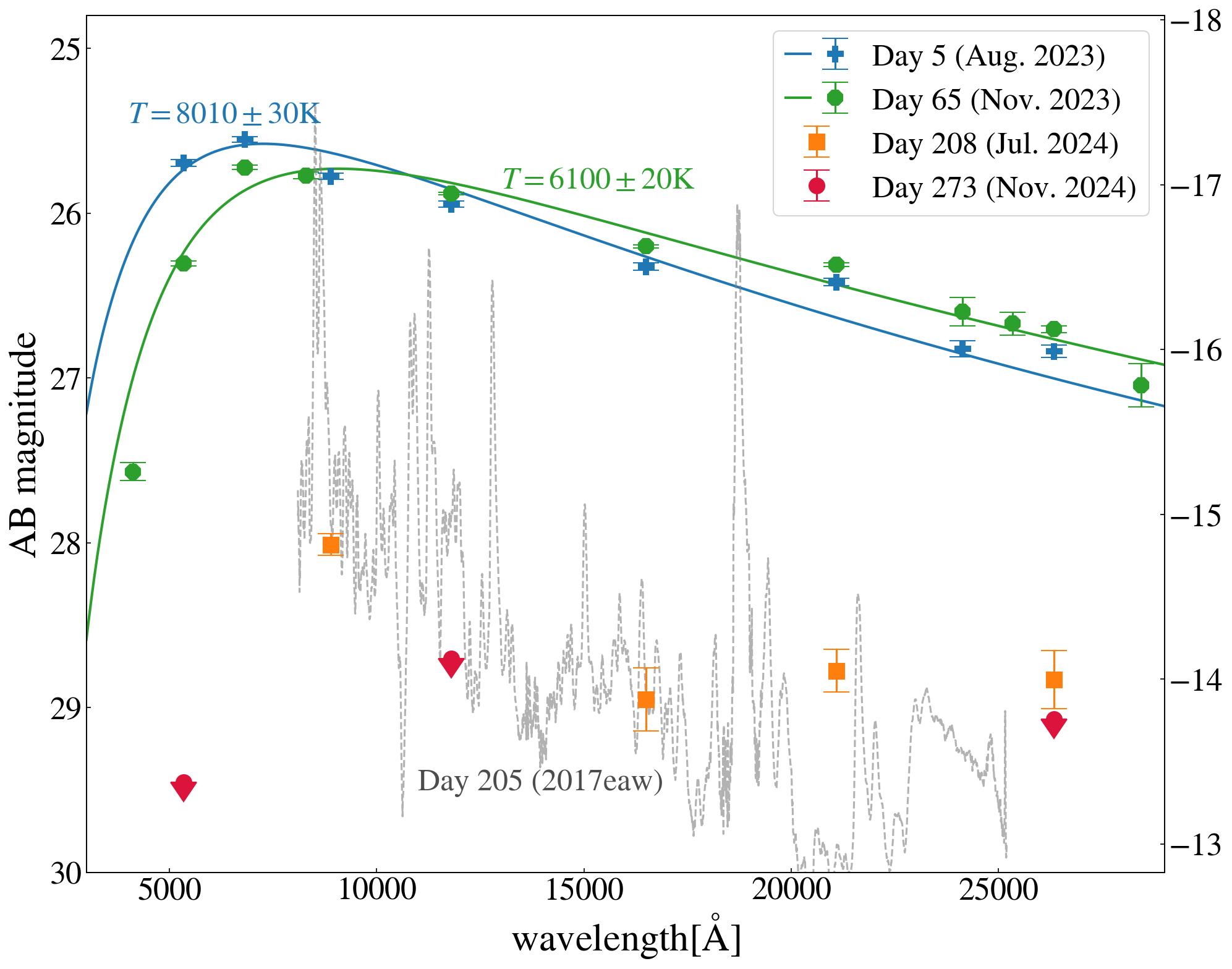}
\caption{Multi-epoch SED fitting results for PLATE-23a. Colored dots represent the observed photometry. For the first and second epochs, for which the SEDs are well described by a blackbody, the corresponding solid lines show the best-fit blackbody models with temperatures of $8010\pm30 \rm~ K$ and $6100\pm20 \rm~ K$, respectively. For the third epoch, which deviates from the optically thick assumption, we plot the best-scaled spectrum of another Type IIP SN (SN 2017eaw, \citet{2018ApJ...864L..20R}) at a similar rest-frame phase (gray dashed line).}
\label{fig:blackbody}
\end{figure}

\begin{figure}[H]
\centering
\includegraphics[width=0.48\textwidth]{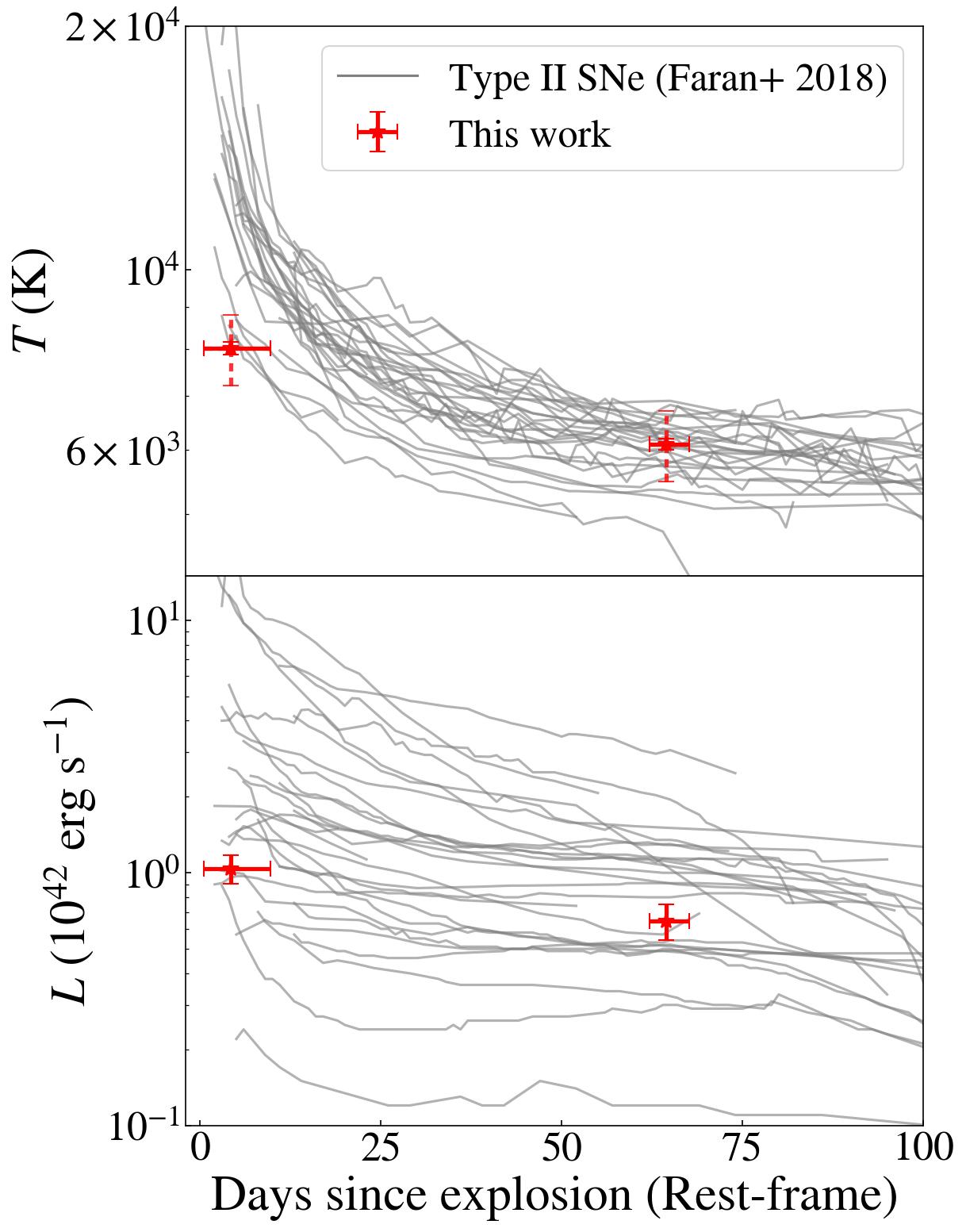}
\caption{Blackbody temperature and bolometric luminosity evolution of PLATE-23a, compared with Type II SNe from \citet{2018MNRAS.473..513F}. The dashed error bars on the blackbody temperature include an additional 10\% relative systematic uncertainty to account for the non-simultaneity of the observations and the impact of emission lines on the photometry.}
\label{fig:LT_compare}
\end{figure}

We begin with the multi-epoch SEDs. As shown in Figure \ref{fig:blackbody}, the first (2023 August) and second (2023 November) epochs are well described by blackbody models, yielding temperatures of $8010\pm30 \rm~ K$ and $6100\pm20 \rm~ K$, and bolometric luminosities of $1.35\pm0.02\times 10^{42} \rm~erg~s^{-1}$ and $8.39\pm0.11\times 10^{41} \rm~erg~s^{-1}$ at rest-frame 5 and 65 days after explosion, respectively. Because the observations within each epoch span a few days across different filters, we have included this range in the time-axis error bars; however, we note that this intra-epoch spread may also introduce systematic uncertainties in the derived parameters, particularly at early times. The temporal evolution of bolometric luminosity and temperature, displayed in Figure \ref{fig:LT_compare}, shows that PLATE-23a follows the trend of Type II SNe \cite{2018MNRAS.473..513F}. It is worth noting that the blackbody temperature derived for the first epoch may be subject to systematic uncertainty. As the F070W band was not observed and the H$\alpha$ emission line falls within the F115W band, this may result in an underestimated temperature. Furthermore, we caution that the error estimates in Figure \ref{fig:LT_compare} only account for the statistical uncertainty of the best-fit SN 2006iw template. Systematic differences between the actual supernova and the template may introduce additional uncertainty in the inferred explosion epoch, potentially causing the rest-frame time after explosion to be underestimated.


The third epoch (2024 July) shows a significant deviation from a blackbody in the SED, indicating that the ejecta have entered the radioactive decay phase. We briefly compare the SED with a scaled spectrum of another Type IIP SN (SN 2017eaw) at a similar rest-frame epoch \cite{2018ApJ...864L..20R}. To model the late-time light curve, we follow the previous works by  \citet{2010ApJ...715..833O,2016MNRAS.459.3939V,2021MNRAS.506.4819P,2021ApJ...906...56D}, which describes the magnitude evolution as:

$$m(t) = -\frac{a_0}{1+e^{(t-t_0)/w_0}}+(s_3\times \frac{t}{100~\rm days})+m_0$$ \label{formula:1}
where the first term represents the transition between the plateau and the radioactive tail using a Fermi-Dirac functional form, the second term accounts for the linear radioactive decay, and the last term is the magnitude in the plateau phase.

Due to the limited number of data points in each band, we estimate typical values of the drop depth in transition ($a_0$), while the length of the plateau ($t_0$) and the slope of the transition phase ($w_0$) could not be effectively constrained. Since the snana-2006iw template ends before the transition (except for F070W, which shows the transition feature in its light curve), the magnitude at the end of the template in each band follows $m_{end}\sim m_0-a_0$. Since the third epoch is already on the radioactive tail, we assumed that $(t - t_0)\gg w_0$, the late-time magnitude can be written as:
$$m = m_{end} + a_0 + s_3\times\frac{t-t_{end}}{100~\rm days}$$

\begin{figure}[h]
\centering
\includegraphics[width=0.48\textwidth]{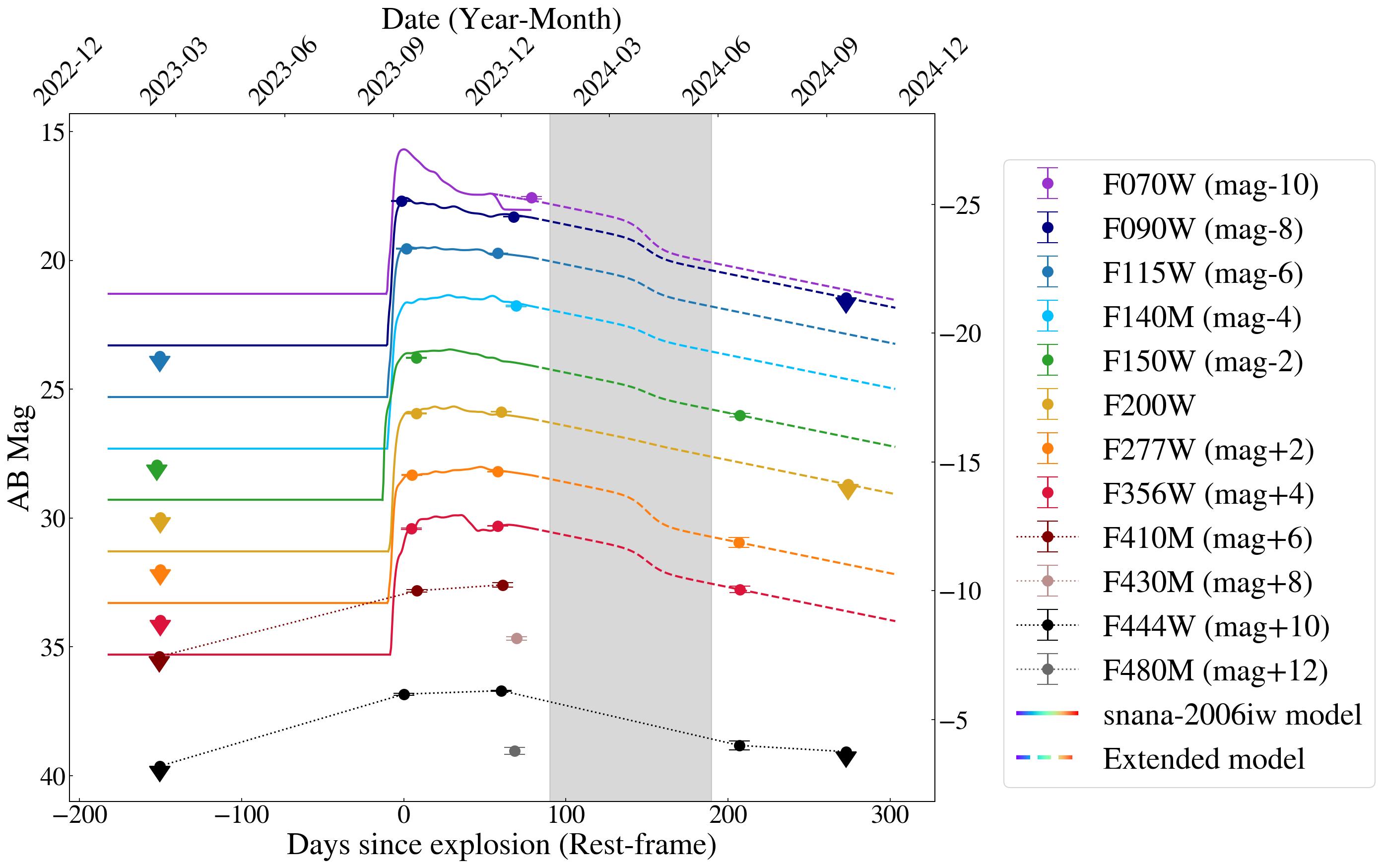}
\caption{Modeling results for the photometric light curve of PLATE-23a. The axes and labels are the same as in Figure \ref{fig:classification}. The light curve before 80 rest-frame days is taken from the snana-2006iw model (the dashed line in the F070W band is the extrapolated light curve, the same as Figure \ref{fig:classification}), while the late-time light curve is modeled following the Formula of \ref{formula:1} in Section \ref{subsec:properties}. The gray-shaded region indicates the uncertainty in $t_0$ and $w_0$ arising from the limited number of photometric data points (we use typical values of $ t_0 = 150~\rm days$ and $w_0 = 5~ \rm days$ in the plot).}
\label{fig:model_lightcurve}
\end{figure}

We adopt a typical value of $s_3=1.28$ (in mag per 100 days), corresponding to the average result for Type II SNe with peak absolute magnitudes similar to PLATE-23a ($-17.6<M_{\rm max}<-17.2$ from \citet{2014ApJ...786...67A}), and estimate $a_0$ in each band as listed in Table \ref{table:stepheight}. We also estimated $a_0$ for the well-studied Type IIP SN 2017eaw in comparable bands: using the LCO V-band light curve from \citet{2019ApJ...876...19S} for F090W (similar rest-frame wavelength coverage) and synthetic photometry of GNIRS spectra from \citet{2018ApJ...864L..20R} for F150W, F200W, F277W, and F356W, respectively. The GNIRS spectra were flux-calibrated using the J, H, Ks light-curve from \citet{2019ApJ...873..127T} (with systematic uncertainty includes in the error of Table \ref{table:stepheight}). The resulting $a_0$ values for SN 2017eaw are generally consistent with those of PLATE-23a, further indicating that this higher-redshift event follows similar late-time evolution as local Type IIP SNe. The final corresponding light curves are shown in Figure \ref{fig:model_lightcurve}. We note that $s_3=1.28$ may have systematic uncertainties owing to differences in the physical properties of the ejecta and the surrounding material of SNe \cite{2014ApJ...786...67A, 2024ApJ...975..132S}. 

\begin{table}[t]
\centering
\footnotesize
\caption{Typical value of the step height between the plateau and radioactive phases of PLATE-23a.}
\label{table:stepheight}
\tabcolsep 12pt 
\begin{tabular*}{0.4\textwidth}{ccc}
 \toprule
 Filter & $a_0$ (PLATE-23a) & $a_0$ (2017eaw) \\
  & mag & mag
    \\
    \hline
    F090W & 0.63(l) & $1.70\pm0.04$ \\
    F150W & $0.28\pm0.07$ & $0.46\pm0.30$ \\
    F200W & 0.05(l) & $0.64\pm0.09$ \\
    F277W & $0.97\pm0.19$ & $0.79\pm0.16$ \\
    F356W & $0.73\pm0.13$ & $0.71\pm0.08$ \\
 \bottomrule
\end{tabular*}
\begin{tablenotes}
\item[1] The $a_0$ value of PLATE-23a is under the assumption of $s_3=1.28$.
\item[2] Letter ``l'' in correspond to 3$\sigma$ lower limit of $a_0$ due to the upper limit in photometry results.
\item[3] The $a_0$ value of 2017eaw is eastimated from LCO V band photometry lightcurve \cite{2019ApJ...876...19S} and GNIRS IR spectra \cite{2018ApJ...864L..20R}.
\end{tablenotes}
\end{table}

\section{Summary} \label{sec:Summary}

In this work, we present the first results of the Pandora Lensing, AGN, and Transient Exploration (PLATE). Using multi-epoch imaging of the Abell 2744 field, we detected a supernova (PLATE-23a) at (03h33m09.9216s, -30d24m01.2229). Our results are summarized as follows:

\begin{itemize}
    \item [1.] With the \texttt{HOTPANTS} difference-imaging technique, we successfully captured the rising and declining phases of PLATE-23a. Further inspection reveals its presence in five distinct epochs of multi-band NIRCam imaging.
    \item [2.] The host galaxy of PLATE-23a has a ground-based spectroscopic redshift of $z=0.688$, placing the transient at a projected distance of $\sim13\rm~kpc$ from the galaxy center. GALFITM modeling yields a Sérsic index $\sim1$. Spatial-resolved SED fitting gives a total stellar mass of $\log M_{*}/M_{\odot}=10.4\pm0.3$, a SFR of $5.1\pm0.1M_{\odot}\rm yr^{-1}$, and inside-out growth features, indicating that the host is a normal star-forming spiral galaxy.
    \item [3.] We classified the multi-band light curve of PLATE-23a using the \texttt{STARDUST2} Bayesian framework. The results strongly favor a Type II(P) SN classification, with snana-2006iw providing the best-fit template. The $M_{\rm max}$ of PLATE-23a is located on the empirical relation between plateau decline rate ($s_2$) and peak magnitude established for local Type II SNe.
    \item [4.] Blackbody fits to the SEDs yield temperatures of $8010\pm30 \rm~ K$ and $6100\pm20 \rm~ K$ and bolometric luminosities of $1.35\pm0.02\times 10^{42} \rm~erg~s^{-1}$ and $8.39\pm0.11\times 10^{41} \rm~erg~s^{-1}$ at rest-frame 5 and 65 days after explosion, consistent with the trend of Type II SNe. The late-time SED is consistent with entering the radioactive decay phase. We estimate the plateau-to-tail drop depth at about $0.3-1.0~\rm mag$ through different bands.
\end{itemize}

These findings demonstrate that multi-epoch deep JWST observations are well-suited for detecting transients and variable sources, particularly at higher redshifts and longer wavelengths. In future work, we will present full analyses of other samples discovered by PLATE, including lensed stars and variability-selected AGNs. We also expect follow-up JWST NIRSpec spectroscopy of selected sources to further constrain their physical properties. Future wide-field surveys such as \textit{Euclid} and \textit{Roman}, combining with JWST instruments, will be able to detect more high-redshift transient events.

\Acknowledgements{
This work is supported by the National Key R\&D Program of China No.2025YFF0510603, the China Manned Space Program with grant no. CMS-CSST-2025-A06, the National Natural Science Foundation of China (grant 12373009), the CAS Project for Young Scientists in Basic Research Grant No. YSBR-062, and the Fundamental Research Funds for the Central Universities. XW acknowledges the support by the Xiaomi Young Talents Program, and the work carried out, in part, at the Swinburne University of Technology, sponsored by the ACAMAR visiting fellowship. 
This work is based on observations made with the NASA/ESA/CSA
James Webb Space Telescope. The data were obtained from the
Mikulski Archive for Space Telescopes at the Space Telescope Science Institute, which is operated by the Association of Universities for Research in Astronomy, Inc., under NASA contract NAS 5-03127 for JWST.
We sincerely acknowledge Macarena Garcia Marin, Stacey Bright, and all the other organizers of the 2025 STScI summer school on high-$z$ transients for their dedication and outstanding efforts.
}

\InterestConflict{The authors declare that they have no conflict of interest.}



\bibliographystyle{ref_temp}
\bibliography{ref}

\end{document}